\documentclass{elsart}

\usepackage{graphicx}
\usepackage{amsmath} 
\usepackage{amsfonts}
\usepackage{amssymb}
\usepackage{setspace}

\usepackage{amssymb}

\begin{document}

\begin{frontmatter}
\journal{Astroparticle Physics}



\title{A Likelihood Method for Measuring the Ultrahigh Energy Cosmic Ray Composition}


\author[Utah]{R.U.~Abbasi}
\author[Utah]{T.~Abu-Zayyad}
\author[LANL]{J.F.~Amman}
\author[Utah]{G.C.~Archbold}
\author[Utah]{K.~Belov}
\author[Utah]{S.A.~Blake}
\author[Montana]{J.W.~Belz}
\author[Columbia]{S.~BenZvi}
\author[Rutgers]{D.R.~Bergman}
\author[Columbia]{J.H.~Boyer}
\author[Utah]{G.W.~Burt}
\author[Utah]{Z.~Cao}
\author[Columbia]{B.M.~Connolly\thanksref{email}}
\author[Utah]{W.~Deng}
\author[Utah]{Y.~Fedorova}
\author[Utah]{J.~Findlay}
\author[Columbia]{C.B.~Finley}
\author[Utah]{W.F.~Hanlon}
\author[LANL]{C.M.~Hoffman}
\author[LANL]{M.H.~Holzscheiter}
\author[Rutgers]{G.A.~Hughes}
\author[Utah]{P.~H\"{u}ntemeyer}
\author[Utah]{C.C.H.~Jui}
\author[Utah]{K.~Kim}
\author[Montana]{M.A.~Kirn}             
\author[Columbia]{B.C.~Knapp}
\author[Utah]{E.C.~Loh}
\author[Utah]{M.M.~Maestas}
\author[Tokyo]{N.~Manago}             
\author[Columbia]{E.J.~Mannel}
\author[LANL]{L.J.~Marek}
\author[Utah]{K.~Martens}
\author[New Mexico]{J.A.J.~Matthews}
\author[Utah]{J.N.~Matthews}
\author[Columbia]{A.~O'Neill}
\author[LANL]{C.A.~Painter}
\author[Rutgers]{L.~Perera}
\author[Utah]{K.~Reil}
\author[Utah]{R.~Riehle}
\author[New Mexico]{M.~Roberts}
\author[Utah]{D.~Rodriguez}
\author[Tokyo]{M.~Sasaki}
\author[Rutgers]{S.~Schnetzer}
\author[Columbia]{M.~Seman}
\author[LANL]{G.~Sinnis}
\author[Utah]{J.D.~Smith}
\author[Utah]{R.~Snow}
\author[Utah]{P.~Sokolsky}
\author[Utah]{R.W.~Springer}
\author[Utah]{B.T.~Stokes}
\author[Utah]{J.R.~Thomas}
\author[Utah]{S.B.~Thomas}
\author[Rutgers]{G.B.~Thomson}
\author[LANL]{D.~Tupa}
\author[Columbia]{S.~Westerhoff}
\author[Utah]{L.R.~Wiencke}
\author[Rutgers]{A.~Zech}
                                               
\address[Utah]{University of Utah, Department of Physics and High
Energy Astrophysics Institute, Salt Lake City, Utah, USA}
\address[LANL]{Los Alamos National Laboratory, Los Alamos, NM, USA}
\address[Adelaide]{University of Adelaide, Department of Physics,
Adelaide, South Australia, Australia}
\address[Montana]{University of Montana, Department of Physics and
Astronomy, Missoula, Montana, USA}
\address[Rutgers]{Rutgers - The State University of New Jersey,
Department of Physics and Astronomy, Piscataway, New Jersey, USA}
\address[Columbia]{Columbia University, Department of Physics and
Nevis Laboratory, New York, New York, USA}
\address[New Mexico]{University of New Mexico, Department of Physics
and Astronomy, Albuquerque, New Mexico, USA}
\address[Tokyo]{University of Tokyo, Institute for Cosmic Ray
Research, Kashiwa, Japan}
\collaboration{The High Resolution Fly's Eye Collaboration}

\thanks[email]{Corresponding author, E-mail:
\texttt{connolly@nevis.columbia.edu}}

\date{\today}

\begin{abstract}
Air fluorescence detectors traditionally determine the dominant chemical composition
of the ultrahigh energy cosmic ray flux by comparing the averaged slant depth of
the shower maximum, $X_{max}$, as a function of energy to the slant depths expected for various 
hypothesized primaries.  In this paper, we present a method to make a direct
measurement of the expected mean number of protons and iron by comparing the shapes 
of the expected $X_{max}$ distributions to the distribution for data. 
The advantages of this method includes the use of information of the full 
distribution and its ability to calculate a flux for various cosmic ray compositions.
The same method can be expanded to marginalize uncertainties due to choice of 
spectra, hadronic models and atmospheric parameters.  We demonstrate the technique 
with independent simulated data samples from a parent sample of protons and 
iron.  We accurately predict the number of protons and iron in the parent sample
and show that the uncertainties are meaningful.
\end{abstract}


\end{frontmatter}

\section{Introduction}

\label{introduction}

Measurements of the energy spectrum, composition and arrival direction 
distributions
elicit clues to the origin of cosmic rays.  
Since a direct measurement and identification of the cosmic ray 
primaries is not possible at these energies, we depend on indirect methods to 
understand the chemical composition of the cosmic ray flux.
The methods vary with the detector type.  
In ground array-type cosmic ray detectors, composition analyses typically focus on the energy-dependence 
of the muon to charged particle ratio, which is thought
to signify the change in primary particle composition as a function of energy\,\cite{agasa_comp}.
Air fluorescence experiments, on the other hand, image the shower development in
the atmosphere and obtain information about the primary composition from $X_{max}$, the slant depth at which the
particle count in the cosmic ray air shower is a maximum.

Showers induced by protons tend to reach the maximum of the shower development at larger 
slant depths than showers induced by heavier elements.  
Since shower-to-shower fluctuations are large, a determination of the primary type for individual showers is not possible, but
averaged quantities like the so-called elongation rate have been successfully used to 
understand the composition of the cosmic ray flux on a statistical basis.  
The elongation rate is defined as the slope of the mean 
height of shower maximum measured in units of 
slant depth, $<X_{max}>$, versus the logarithm of 
the primary cosmic ray energy observed.  The measured elongation rate is
compared to that of simulated proton and iron primaries to assess whether
the cosmic ray composition is proton-like or heavier\,\cite{bird1993,proto,archbold}.   

\begin{figure}[t]
\includegraphics[width=1.\textwidth]{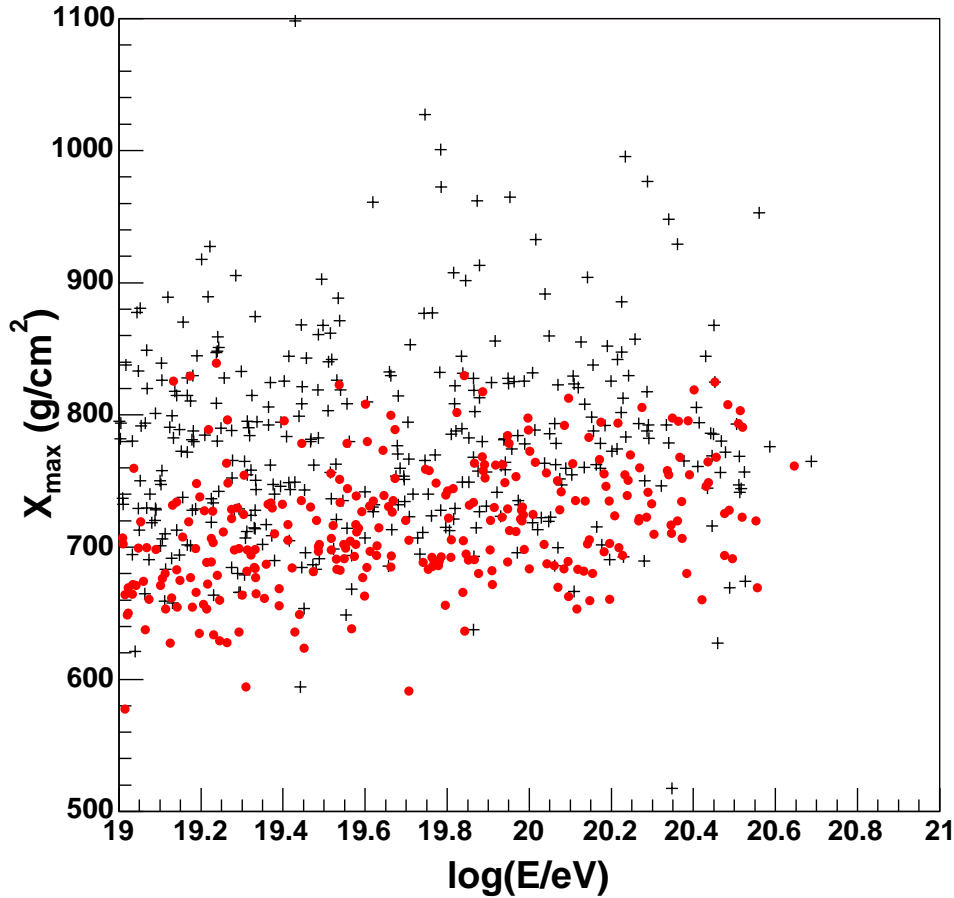}
 \caption{{\it \label{fig:elongation} Simulated cosmic ray events show 
 shower-to-shower fluctuations which cause a large overlap between
 proton (black) and iron (red) events when plotted as a function of $X_{max}$ and $log(E/eV)$.
 The QGSJet hadronic model was used under the CORSIKA extensive air shower simulator
 to model the shower profiles.  These events were then passed through a simulation of the 
 HiRes stereo detector response.}}
\vskip 1 cm 
\end{figure}

Due to the relative insensitivity of ultrahigh energy cosmic ray detectors to small shifts 
in composition between various low-Z or high-Z primaries,
their observations have historically been compared to what is expected for 
``protons'' and ``iron,''
with the two elements as stand-ins for all light and heavy elements.
For $<X_{max}>$ (see Section\,\ref{section:demonstrating}),
we expect $\sim 100\,\rm{g/cm^2}$ difference between protons and iron.  However, 
as shown in Fig.\,\ref{fig:elongation}, large shower-to-shower
fluctuations and the detector resolution cause a significant overlap in their distributions
($\sim 40-70\,\rm{g/cm^2}$). 

Here, we present a method to answer the question: 
in a fluorescence data sample where one has
measured $X_{max}$ for each event, 
what is the best estimate for the expected 
mean\,\footnote{Here we will refer to the mean although we are taking a Bayesian approach.
The number of protons, iron, etc. predicted by the likelihood can better 
understood as the best estimate of the ``true" number of 
protons and iron given the universe's cosmic
ray flux 
and the HiRes aperature rather than the mean number expected for an infinite 
number of HiRes experiments.  
(Here the universe's cosmic ray flux is assumed to be homogeneous 
and isotropic as 
there is currently no evidence to the contrary.  However, 
there is nothing to prevent us binning $X_{max}$ as a function
of sky coordinates.)  
Regardless, for brevity, we use frequentist language.}   
number of protons and iron
from that experiment?  This question is not directly answered through other methods.
Once the answer to this question is obtained, the mean number can be converted to 
a flux for protons and iron.
This method can be expanded to evaluate the contribution of any chemical element.
Further, this method uses more information than averaged 
quantities like the elongation rate; it makes use 
of the shape of the $X_{max}$ distribution rather than simply its mean.
Lastly, the technique allows the introduction of
systematic uncertainties and  prior knowledge or lack thereof concerning the
composition spectra, hadronic generators, parameters concerning the 
extensive air shower and atmospheric uncertainties.

In brief, the technique for extracting the cosmic ray composition essentially requires us to 
find the best normalizations for the ``hypothesized'' $X_{max}$ distributions for protons and iron.
By ``best'' we mean the sum of the expected mean events for the 
hypothesized distributions that best match the expected mean for the
$X_{max}$ distribution for data.
The chance that the combined hypothesized distributions match 
the parent distribution for data
is proportional to the likelihood, $l(D|p,f)$.
The likelihood $l(D|p,f)$ is maximized for the best estimate for 
the expected mean number of protons ($p$) and iron ($f$)
where $D$ denotes the data sample.
The variables $p$ and $f$ sum to the expected mean number of data events, $p+f=d$.
Note the expected means for protons and iron add to the expected mean number of events
and not to the total events in the given data sample.  

We develop the method using simulated stereoscopic data
for the High Resolution Fly's Eye (HiRes) experiment in Utah.  The method is 
tested with many independent data sets that originate from a simulated 
parent data set with $p_{true}$ proton events generated with an $E^{-\alpha_p}$ spectrum
and $f_{true}$ iron events generated with an $E^{-\alpha_f}$ spectrum. 
Each bin of this parent distribution is fluctuated according to Poisson statistics 
to build a set of independent simulated data distributions with which we test the method of 
predicting $p$ and $f$.  
A general 
description of this technique is given in\,\cite{bhat}.
To include typical instrumental uncertainties in our study, we will
apply the method to a particular cosmic ray experiment, the HiRes stereo 
experiment.  All simulated showers used in this paper have been subjected to the 
full HiRes stereo simulation and reconstruction.

The paper is organized as follows:  Section\,\ref{section:detector} 
gives an overview of the HiRes detector, which we use as the detector in the simulated data.  
Section\,\ref{section:likelihood} gives a  description of the likelihood method used to find the most likely mixture of
protons and iron in the simulated data.  In Section\,\ref{section:demonstrating}, the likelihood method is first 
modified to account for our ignorance of the proton and iron spectra, and then demonstrated with
simulated data samples.  Finally, Section\,\ref{section:conclusion} states conclusions and results. 

\section{The HiRes Detector}
\label{section:detector}
HiRes is an air fluorescence experiment with two sites
(HiRes\,1\,\&\,2) at the US Army Dugway Proving Ground in the
Utah desert ($112^{\circ}$\,W longitude, $40^{\circ}$\,N latitude,
vertical atmospheric depth $860\,{\mathrm{g}}/{\mathrm{cm}}^{2}$).
The two sites are separated by a distance of 12.6\,km.

Each of the two HiRes ``eyes'' comprises several telescope units
monitoring different parts of the night sky.  With 22 (42)
telescopes with 256 photomultiplier tubes each at the first (second)
site, the full detector covers about $360^{\circ}$ ($336^{\circ}$)
in azimuth and $3^{\circ}-16.5^{\circ}$ ($3^{\circ}-30^{\circ}$) in
elevation above horizon.  Each telescope consists of a mirror with an 
area of about $5\,\mathrm{m}^{2}$ for light collection and a 
cluster of photomultiplier tubes in the focal plane.

A cosmic ray primary interacting in the upper atmosphere induces an
extensive air shower which the detectors observe as it develops in
the lower atmosphere.  
The photomultiplier tubes triggered by the shower define an
arc on the sky, and, together with the position of the detector, the
arc determines the so-called shower-detector plane.  When an air shower
is observed in stereo, the shower trajectory is in principle simply the
intersection of the two planes.  This method can be further improved by also
taking advantage of the timing information of the tubes, and in our
analysis the shower geometry is determined by a global $\chi^2$
minimization using both the timing and pointing information of all tubes.

The next step in the 
reconstruction is to calculate the shower development profile. 
However, light arriving at the detector  is collected by discrete 
PMTs, each of which covers about  $1^{\circ}\times 1^{\circ}$ of the sky.  
The signal from a longitudinal segment of the EAS is thus necessarily split among many PMTs.  
For profile fitting, the signal must be re-combined into bins that correspond to the 
longitudinal segments of the EAS in HiRes 1 and HiRes 2. 
The re-binned signal, corrected for atmospheric 
extinction and with \v{C}erenkov light subtracted is fit to a Gaisser-Hillas functional form 
(Eq. \ref{gh}) using a $\chi^2$ that fits the function to the
profiles measured in the two detectors. This form has been shown to be in good agreement with EAS simulations 
\cite{101,89,87} and with HiRes data \cite{102}.

\begin{equation}
N(X)=N_{max}{\left(\frac{X-X_{\circ}}{X_{max}-X_{\circ}}\right)}^
{(X_{max}-X_{\circ})/\lambda}\exp\left[\frac{(X_{max}-X)}{\lambda}\right].
\label{gh}
\end{equation}

From measurements of laser tracks and stars in the field of view
of the cameras we estimate that the systematic error in the arrival
direction determination is not larger than $0.2^{\circ}$, mainly caused
by uncertainties in the survey of mirror pointing directions.

Various aspects of the HiRes detector and the 
reconstruction procedures are described in \cite{nim2002,star2002,matthews2003}.

\section{The Likelihood}
\label{section:likelihood}
To calculate the best estimate of $p$ and $f$, the mean number of protons 
and iron expected (at a HiRes experiment), we maximize $l(D|p,f)$, 
the likelihood of the data sample $D$ given $p$ and $f$.
To this end, Bayes' theorem allows us to convert the probability of a 
particular data sample $D$ given the composition mixture, $P(D|p,f)$, to the 
probability of the composition mixture given the data, $P(p,f|D)$:
\begin{equation}
\label{equation:bayes}
P(p,f|D) = \frac{P(D|p,f)q(p,f)}{\sum_{p,f}P(D|p,f)q(p,f)}.
\end{equation}

In this expression, $q(p,f)$ is the prior probability of $p$ and $f$.  
Initially, the prior probability will contain information about the 
expected distributions for protons and iron and their respective energy 
spectra.  However, in principle, it can contain any number
of (un)knowns concerning the hadronic generator, uncertainties in
atmospheric parameters, etc.  We effectively maximize $P(p,f|D)$ by maximizing 
$l(D|p,f)\equiv P(D|p,f)q(p,f)$.

To calculate $l(D|p,f)$, we divide the $X_{max}$ distributions into $N$ bins. 
We let $P_i$ and $F_i$ be the number of events in the $i^{th}$ bin
of the simulated proton and iron samples, respectively, and $D_i$ be
the number of data events in the $i^{th}$ bin.  (The capitalized quantities
are ones which we know at the outset of the calculation.)  

While we typically generate samples of simulated events that are much larger 
than the data sample, the sets $P_i$ and $F_i$ nevertheless represent single
instances or fluctuations around the true, unknown means $\tilde{p}_i$ and
$\tilde{f}_i$ for the simulated samples.  Similarly, the number of data
events $D_i$ in the $i^{th}$ bin fluctuates around the true mean count 
$\tilde{d}_i$.  Assigning a Poisson probability to the counts in each bin
$D_i$, $P_i$, and $F_i$, we can write the likelihood function as:

\begin{equation}
\label{equation:dpf_likelihood}
l(D|\{\tilde{d}_i\},\{\tilde{p}_i\},\{\tilde{f}_i\}) = \prod_{i=1}^N 
           \left( \frac{(\tilde{d}_i)^{D_{i}}e^{-\tilde{d}_i}}{D_{i}!} \right) 
           \left( \frac{(\tilde{p}_i)^{P_{i}}e^{-\tilde{p}_i}}{P_{i}!} \right)
           \left( \frac{(\tilde{f}_i)^{F_{i}}e^{-\tilde{f}_i}}{F_{i}!} \right)
\end{equation}

The actual values for $\tilde{p}_i$ and $\tilde{f}_i$ will not interest us;
these are nuisance parameters which we will eventually eliminate.  However,
the mean number of data counts expected in the $i^{th}$ bin is expressed
as a weighted sum of these parameters:

\begin{equation}
\label{equation:mean_d}
\tilde{d}_i = \epsilon_p\tilde{p}_i+\epsilon_f\tilde{f}_i.
\end{equation}

The purpose of the weights $\epsilon_p$ and $\epsilon_f$ is to 
1) scale down the (presumably larger) simulated sample sizes to the 
data sample size, and
2) set the relative mixture of protons and iron expected in the data.  Hence,
the quantities which we want to estimate are $\epsilon_p$ and $\epsilon_f$.
Inserting Eq.\,(\ref{equation:mean_d}) into Eq.\,(\ref{equation:dpf_likelihood})
and marginalizing (integrating out) the nuisance parameters 
$\tilde{p}_i$ and $\tilde{f}_i$ we define the global likelihood function
$l(D|\epsilon_p,\epsilon_f)$ as:
\begin{eqnarray}
\label{equation:likelihood_special}
l(D|\epsilon_p,\epsilon_f) = \prod_{i=1}^{N~bins}
\int d\tilde{p}_i \int d\tilde{f}_i & 
  \left( \frac{(\epsilon_p\tilde{p}_i+\epsilon_f\tilde{f}_i)^{D_i}
          e^{-(\epsilon_p\tilde{p}_i+\epsilon_f\tilde{f}_i)}}{D_i!} \right)
\nonumber\\
& \times \left( \frac{(\tilde{p}_i)^{P_{i}}e^{-\tilde{p}_i}}{P_{i}!} \right)
         \left( \frac{(\tilde{f}_i)^{F_{i}}e^{-\tilde{f}_i}}{F_{i}!} \right).
\end{eqnarray}

In fact, the integration can be performed exactly 
(see \cite{bhat} for details), and reduces to:
\begin{eqnarray}
l(D|\epsilon_p,\epsilon_f) = \prod_{i=1}^N \sum_{n=0}^{D_i} &
    \binom{P_i+D_i-n}{D_i-n} 
    \frac{(\epsilon_p)^{D_i-n}}{(1+\epsilon_p)^{P_i+D_i-n+1}}
\nonumber\\
& \times 
    \binom{F_i+n}{n}
    \frac{(\epsilon_f)^n}{(1+\epsilon_f)^{F_i+n+1}}.
\end{eqnarray}

In practice, then, the function we maximize is not $l(D|p,f)$ directly;
rather, we maximize $l(D|\epsilon_p,\epsilon_f)$ (or, more
precisely, we minimize $-\log l(D|\epsilon_p,\epsilon_f)$) 
with respect to both $\epsilon_p$ and $\epsilon_f$.

With our best estimates for $\epsilon_p$ and $\epsilon_f$,
we can now estimate the mean number of proton and iron events $p$ and $f$
expected for the experiment.  The estimates suggested in~\cite{bhat} are
\begin{equation}
\label{equation:protons}
p=\epsilon_p (\sum_{i=1}^{N} P_i +N)
\end{equation}
and
\begin{equation}
\label{equation:iron}
f=\epsilon_f (\sum_{i=1}^{N} F_i +N)
\end{equation}
which of course reduce to the more obvious estimates 
$p=\epsilon_p \sum P_i$ and $f=\epsilon_f \sum F_i$
in the usual situation when the number of simulated events is much greater
than the number of bins in $X_{max}$.

\section{Demonstrating the Method}
\label{section:demonstrating}
To demonstrate the method, we create many simulated data 
samples from a single parent distribution made 
with $p_{true}$ proton and $f_{true}$ iron simulated events
with energies in excess of $10^{18.5}$\,eV.  
First, we generate a library of showers of various energies
and primary particles with the CORSIKA simulator
using the QGSJET hadronic model.  For a given composition, 
these showers are 
selected randomly according to their energy spectrum 
and then simulated in the detector with a random geometry.  
We expect the energy spectra to have an energy dependence between
$E^{-2}$ and $E^{-3}$.  In our first application 
of the method, we will pick $\alpha_p=2$
and $\alpha_f=3$, a choice motivated by previous
composition analyses\,\cite{proto}.  This will just serve as 
an example; we will drop this assumption in the 
next section, where we show how to eliminate the spectral 
dependence.
The simulated data samples are subject to the same quality 
cuts that we would apply to real data (see e.g.\,\cite{ref9}).  
We require a minimum track length 
of $3^{\circ}$ in each detector, an estimated angular uncertainty
in both azimuth and zenith angle of less than
$2^{\circ}$, and a zenith angle less than $70^{\circ}$.  We additionally 
require an estimated energy uncertainty
of less than $20\%$
and $\chi^2/{\rm dof}<5$ for both the energy and the geometry
fit, 
and that $X_{max}$ appear in the field of view 
one of the detectors.
Lastly, the reconstructed primary particle 
energy must be greater than $10^{19}$\,eV.

\begin{figure}[t]
\includegraphics[width=1.\textwidth]{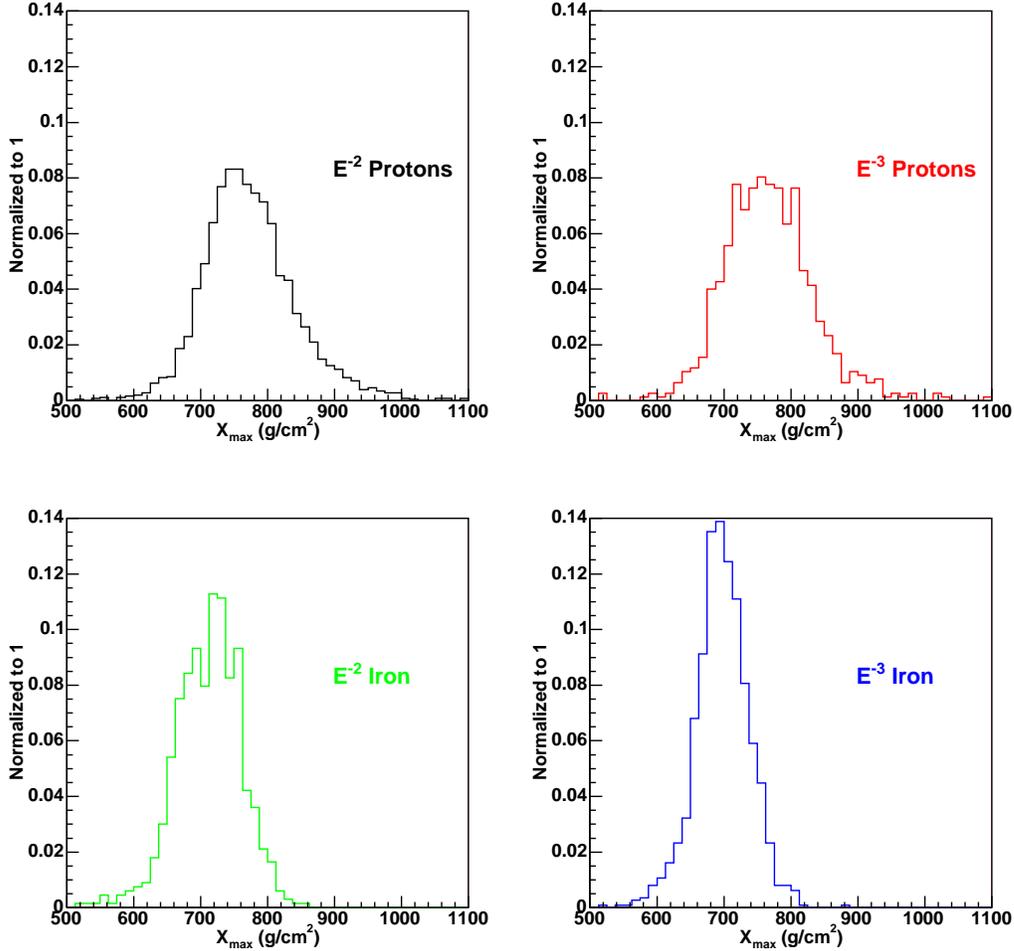}
 \caption{\label{fig:hypotheses} Iron and proton $X_{max}$ 
 distributions generated
 with $E^{-2}$ and $E^{-3}$ spectra with energies above $10^{19}$ eV.}
\end{figure}

After applying the quality cuts to the parent sample, we
fluctuate each bin in the $X_{max}$ distribution 
in the parent data sample many times 
according to Poisson statistics to obtain 
many independent data samples.
For the moment, we assume that we know that
$\alpha_p = 2$ and $\alpha_f =3$.
If the method works, we expect the means for $p_{max}-p_{true}$  
and $f_{max}-f_{true}$ to be 0, and the calculated
uncertainties for $p$ and $f$ should be meaningful; that is,
the quoted values for $p$ and $f$  are within 
$1\,\sigma$ of $p_{true}$ and $f_{true}$ in $68\%$ 
of the data samples.  Figure\,\ref{fig:hypotheses} 
shows the proton and iron distributions used to calculate
the likelihood.  We use the same events for 
generating the parent distribution and for the 
hypothesized distributions in the likelihood.
We generate 2571, 773, 665 and 1121 events for
$E^{-2}$ protons, $E^{-3}$ protons, $E^{-2}$ iron and $E^{-3}$ iron.
A similar but independent set of distributions were 
combined to build the `fake' data sample.

\begin{figure}[t]
\includegraphics[width=1.\textwidth]{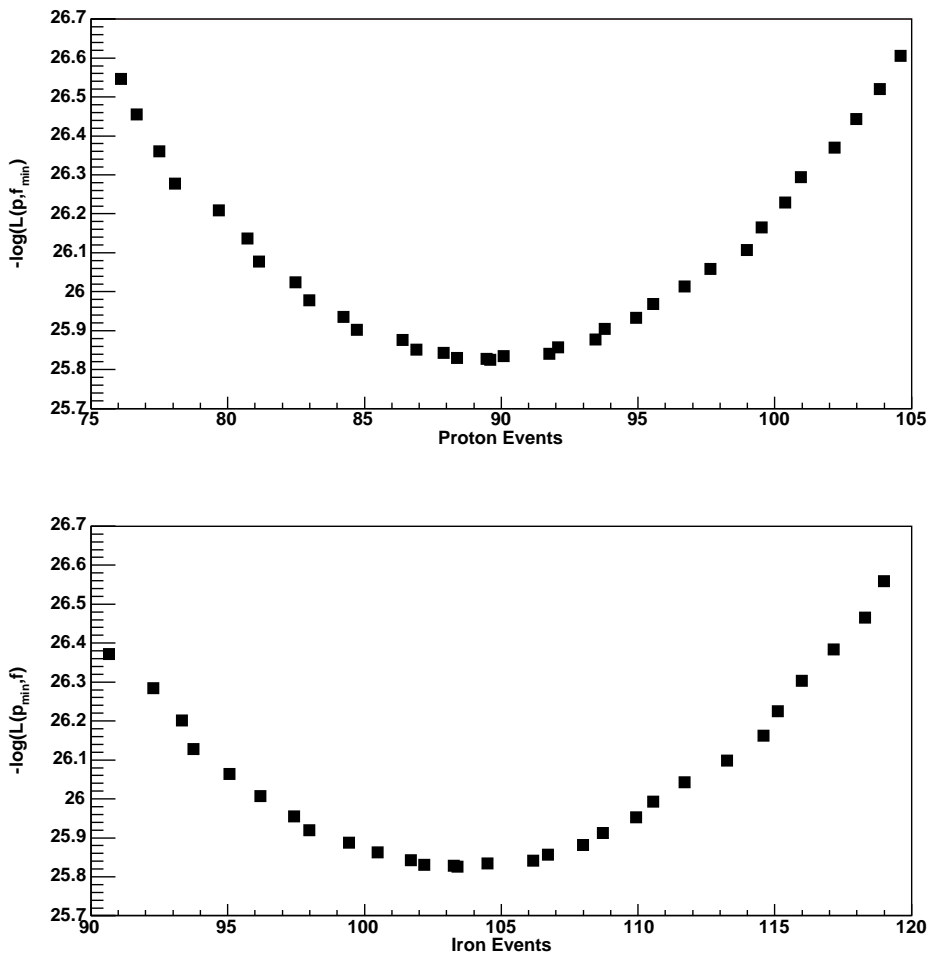}
 \caption{{\it \label{fig:likelihood_200_0.5_0.5_1_0_2} $-\log{l(D|p,f_{max})}$ (top) and 
 $-\log{l(D|p_{max},f)}$ (bottom) evaluated
 for one simulated data sample with 86 proton and 117 iron events
 fluctuated from 100 proton and 100 iron events.  Although regions about the minima 
 are parabolic, the distributions have substantial tails.}}
\vskip 1 cm 
\end{figure}

We first test the method with a mixture of $50\%$ protons and $50\%$ 
iron.  
Figure\,\ref{fig:likelihood_200_0.5_0.5_1_0_2} shows $-\log{l(D|p,f_{max})}$ and $-\log{l(D|p_{max},f)}$,
for one simulated data set with 86 protons and 117 iron events
fluctuated from a parent sample of 100 proton and 100 iron events;
$p_{max}$ ($f_{max}$) is where $l(D|p,f)$ is maximized with repect to $p$ ($f$).
From the parabolic shape of these distributions one might suspect that 
the $1\,\sigma$ uncertainties for $p$ and $f$ can be calculated 
through the equations
\begin{equation}
\label{equation:error}
-\log{ l(p_{max} \pm \sigma_{p},f_{max})} =-\log{ l(p_{max})}
+\frac{1}{2}
\end{equation}
and
\begin{equation}
-\log{l(p_{max},f_{max} \pm \sigma_{f})} =-\log{ l(f_{max})}
+\frac{1}{2}
\end{equation}
where $\sigma_p$ and $\sigma_f$ are the uncertainties for $p$ and $f$, respectively.
However, we find that $l(D|p_{max},f)$ and $l(D|p,f_{max})$ have substantial tails and so their uncertainties
have to be calculated numerically.  

\begin{figure}[t]
\includegraphics[width=1.\textwidth]{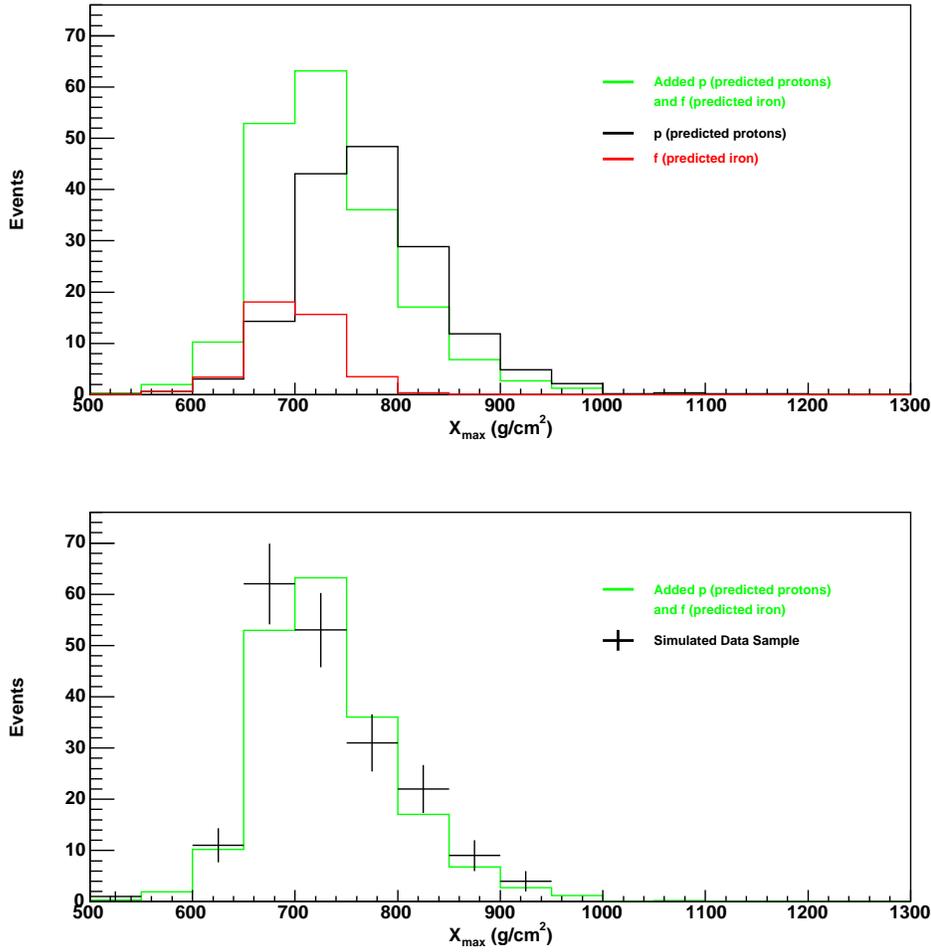}
 \caption{\it \label{fig:example} The distributions for protons and iron normalized
 to the number of predicted events, the total predicted mean, and a simulated data set
 of 86 proton and 117 iron events
 created from a parent distribution of 100 proton and 100 iron events.  
 The $\chi^2$ between the parent data set and the predicted sum is 6.7 with 8 degrees of freedom.}
\vskip 1 cm 
\end{figure}

Figure\,\ref{fig:example} shows the corresponding distributions for protons and iron normalized 
to the number of predicted events, the total predicted mean and the simulated data set.
We calculate the $\chi^2$ for this fit to be 6.7 with 8 degrees of freedom while keeping
in mind that the total events predicted by the sum of $p$ and $f$ are not meant
to predict the total number of events, but rather the expected mean number of events.     

To artificially separate 
the systematic and statistical uncertainties, one can, for example, remove 
any prior knowledge concerning the uncertainties in our models for protons and iron (i.e. remove 
$q(\{p_i\},\{f_i\},\epsilon_pp,\epsilon_ff)$ from \linebreak $l(D|\{p_i\},\{f\},\epsilon_pp,\epsilon_ff)$) 
and define statistical uncertainty as 
\begin{equation}
\label{equation:l}
l_{stat}(D|p,f)\equiv \prod_{i=1}^{N}\int dp_i \int df_i \frac{\tilde{d}_i^{D_i}e^{-\tilde{d}_i}}{D_i!}
\end{equation}
Then, by calculating the total uncertainty ($\sigma_{tot}$) from $l(D|p,f)$
and $\sigma_{stat}$ from $l_{stat}(D|p,f)$, 
the systematic uncertainty ($\sigma_{sys}$)
can be deduced from
\begin{equation}
\sigma_{sys}=\sqrt{\sigma_{tot}^2-\sigma_{stat}^2}.
\end{equation}
Notice we have assumed that the $l_{stat}(D|p,f)$ and $l(D|p,f)$ both 
have maxima at the same $p_{max}$ and $f_{max}$.
This is not the case in general.  However, 
in the instance where the maxima do not occur at the same $p_{max}$ and $f_{max}$,
the difference arises from choosing what is in fact an arbitrary definition of 
the statistical uncertainty, $l_{stat}(D|p,f)$.

\begin{figure}[t]
\includegraphics[width=1.\textwidth]{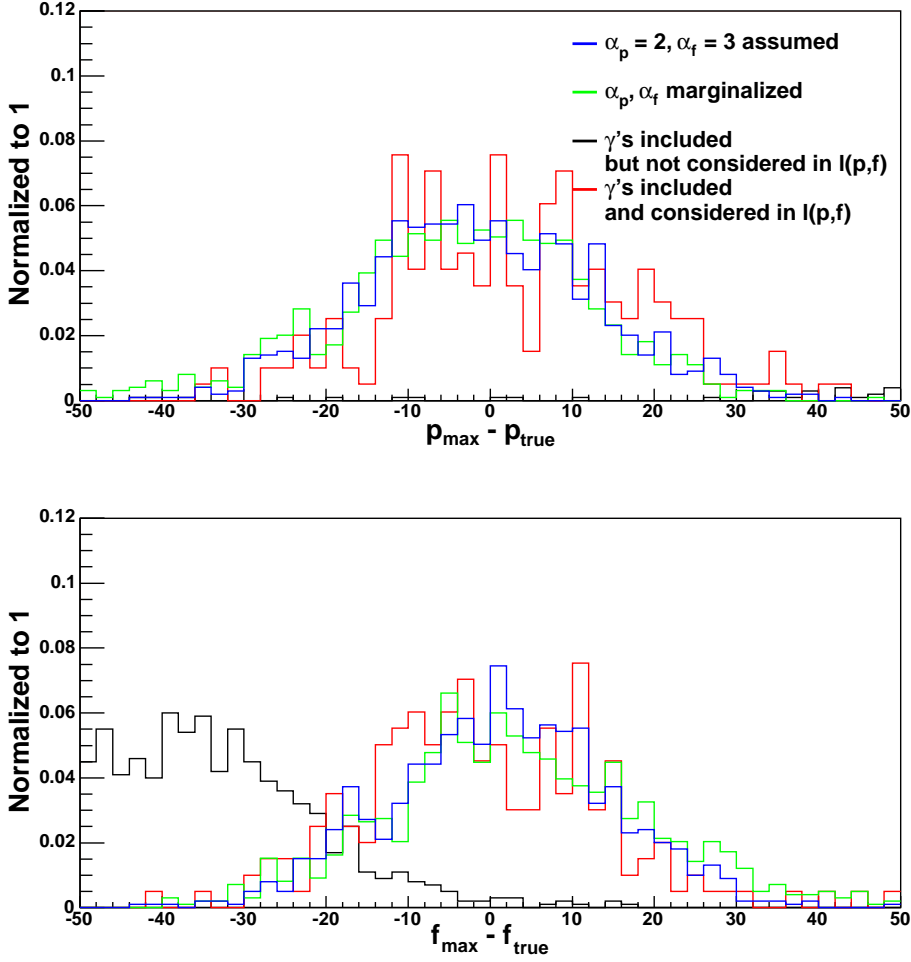}
 \caption{{\it \label{fig:diff} Histograms of $p$-$p_{true}$ (top) and $f$-$f_{true}$ (bottom)
 for $p_{true}=100$ and $f_{true}=100$.  The green (blue) histograms were calculated with 
 Eq.\,(\ref{equation:likelihood_full}) (Eq.\,(\ref{equation:likelihood_special})). 
 The black histogram makes use of Eq.\,(\ref{equation:likelihood_full}) that is applied to
 a parent data sample made of 250 events with 100 proton, 100 iron and 50 gammas.
 The red histogram includes an additional term in Eq.\,(\ref{equation:likelihood_full})
 to account for gammas.  The histograms have overflow events (e.g., most of the events
 in the black histogram lie outside the abscissa limits).} }
\vskip 1 cm 
\end{figure}

\begin{figure}[t]
\includegraphics[width=1.\textwidth]{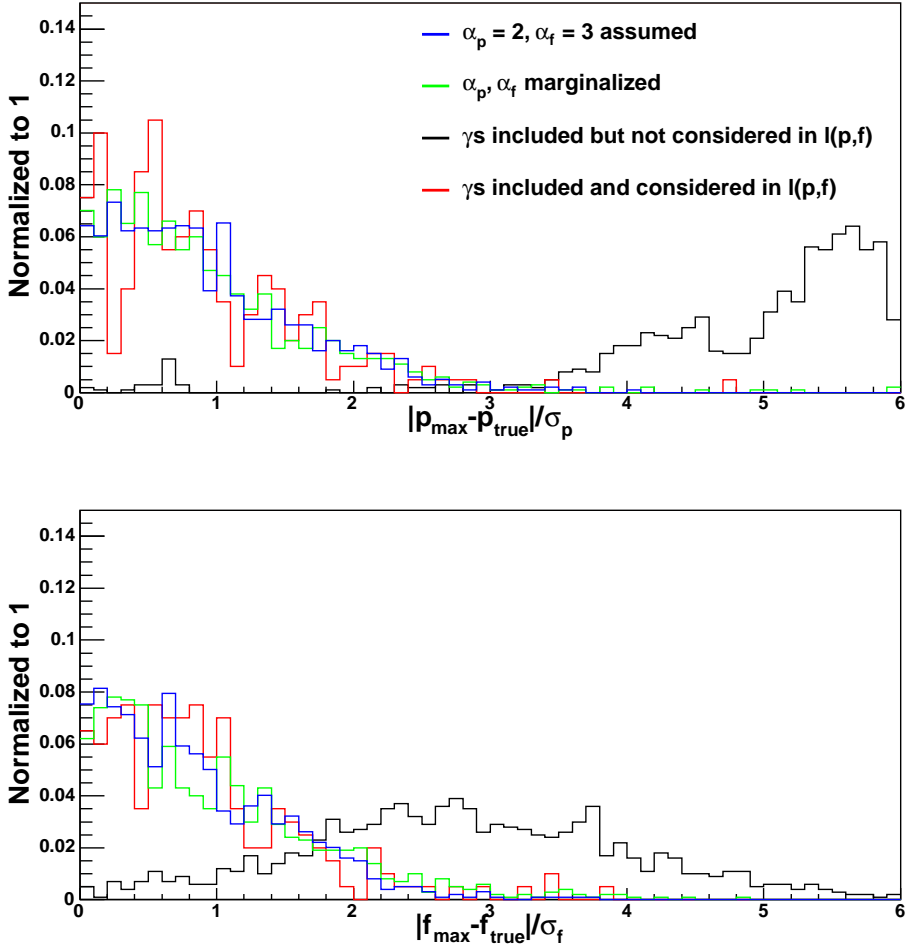}
 \caption{{\it \label{fig:sigma} Deviation from $p_{true}$ and $f_{true}$
 expressed relative to the proton and iron uncertainties, $\sigma_p$ and $\sigma_f$, respectively.  
 The green (blue) histograms were calculated with 
 Eq.\,(\ref{equation:likelihood_full}) (Eq.\,(\ref{equation:likelihood_special})). 
 The black histogram makes use of Eq.\,(\ref{equation:likelihood_full}) that is applied to
 a parent data sample made of 250 events with 100 proton, 100 iron and 50 gammas.
 The red histogram includes an additional term in Eq.\,(\ref{equation:likelihood_full})
 to account for gammas.  The histograms have overflow events (e.g., many of the events
 in the black histogram lie outside the abscissa limits).} }
\vskip 1 cm 
\end{figure}

\begin{figure}[t]
\includegraphics[width=1.\textwidth]{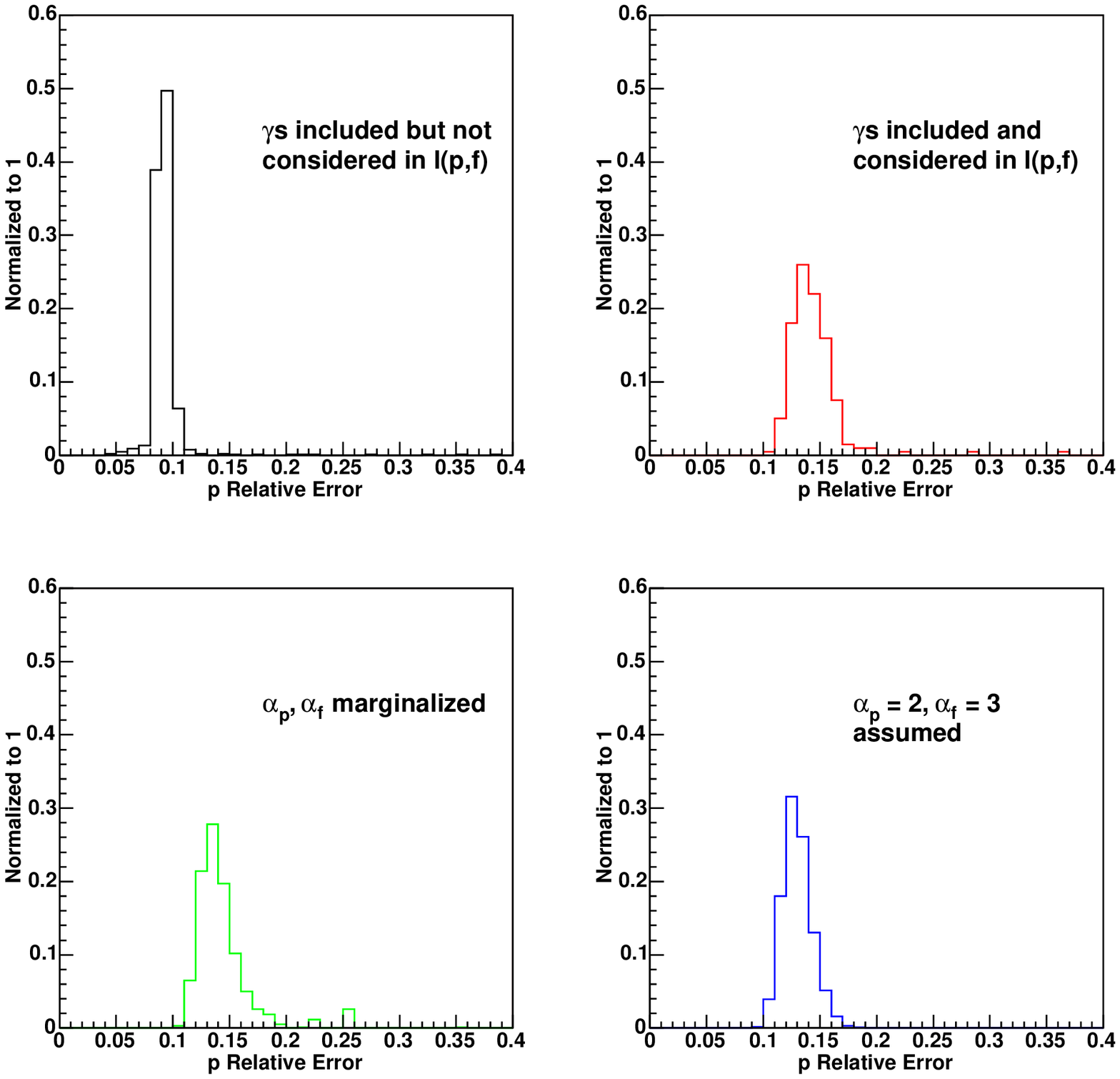}
 \caption{\label{fig:errordiff_p} {\it Fractional uncertainty for the mean number of
 protons ($p$). 
 The green (blue) histograms were calculated with 
 Eq.\,(\ref{equation:likelihood_full}) (Eq.\,(\ref{equation:likelihood_special})). 
 The black histogram makes use of Eq.\,(\ref{equation:likelihood_full}) that is applied to
 a parent data sample made of 250 events with 100 proton, 100 iron and 50 gammas.
 The red histogram includes an additional term in Eq.\,(\ref{equation:likelihood_full})
 to account for gammas.  The histograms have overflow events.} }
\vskip 1 cm 
\end{figure}

\begin{figure}[t]
\includegraphics[width=1.\textwidth]{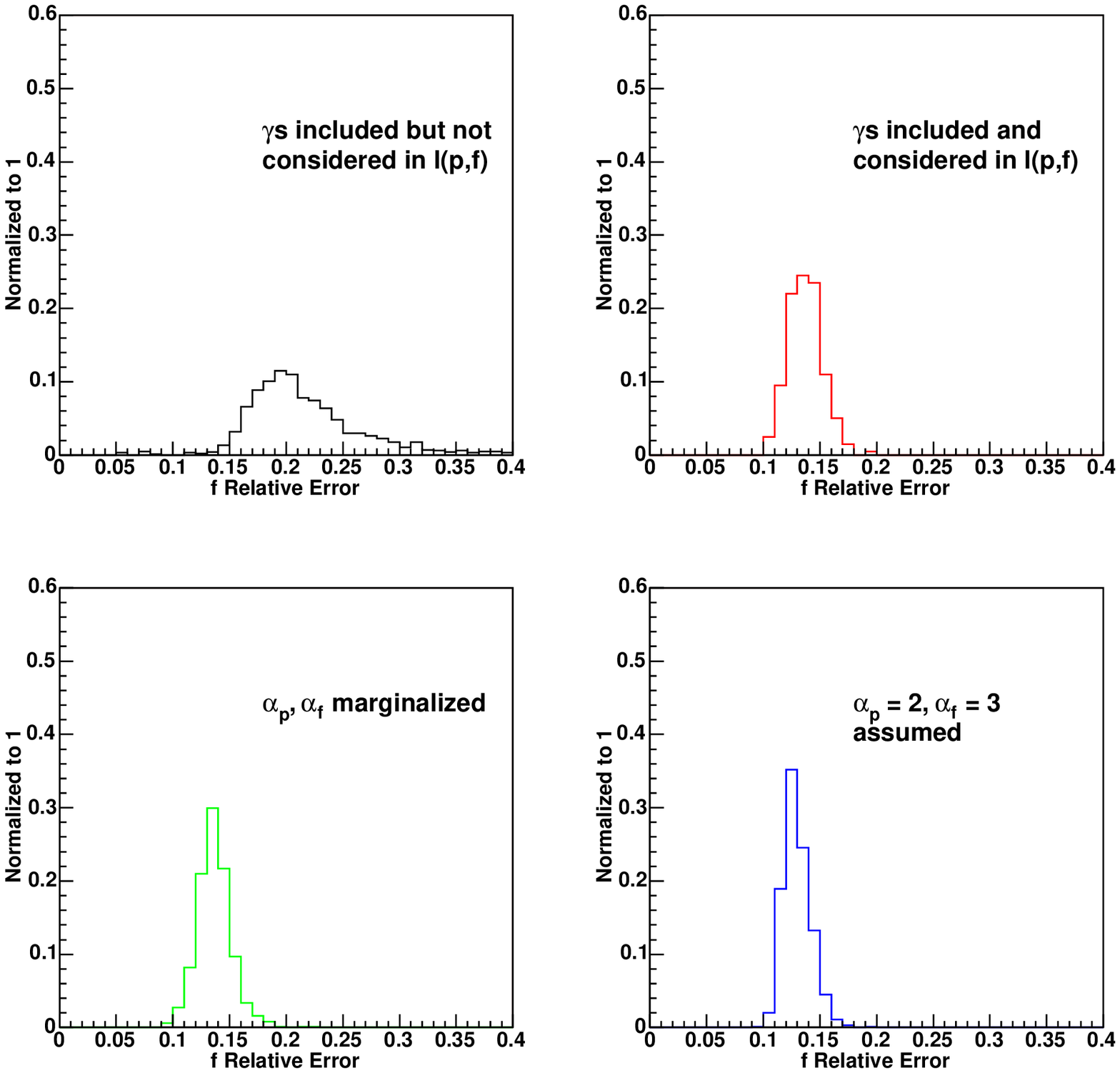}
 \caption{\label{fig:errordiff_f} {\it Fractional uncertainty for the mean number 
 of iron ($f$). 
 The green (blue) histograms were calculated with 
 Eq.\,(\ref{equation:likelihood_full}) (Eq.\,(\ref{equation:likelihood_special})). 
 The black histogram makes use of Eq.\,(\ref{equation:likelihood_full}) that is applied to
 a parent data sample made of 250 events with 100 proton, 100 iron and 50 gammas.
 The red histogram includes an additional term in Eq.\,(\ref{equation:likelihood_full})
 to account for gammas.  The histograms have overflow events.} }
\vskip 1 cm 
\end{figure}

Figure\,\ref{fig:diff} shows the distributions of $p_{max}-p_{true}$ and $f_{max}-f_{true}$
histogrammed in blue.  
The means of these distributions
do not deviate more than $1.3\%$ for protons and iron -
an insignificant deviation considering 
the mean uncertainties in Fig.\,\ref{fig:sigma}.
Figure\,\ref{fig:sigma} shows the distribution of
$|p_{max}-p_{true}|/\sigma_p$ and $|f_{max}-f_{true}|/\sigma_f$ where
$\sigma_p$ and $\sigma_f$ are the uncertainties for $p$ and $f$, respectively.  As expected,
$68\%$ of the predictions for $p$ and $f$ deviate less than $1\sigma$ from the 
$p_{true}$ and $f_{true}$.  Figures\,\ref{fig:errordiff_p} and \ref{fig:errordiff_f}
shows the relative error for protons and iron, respectively.  
The most probable error for $p$ and $f$ is $\sim 13\%$.  


\begin{figure}[t]
\includegraphics[width=1.\textwidth]{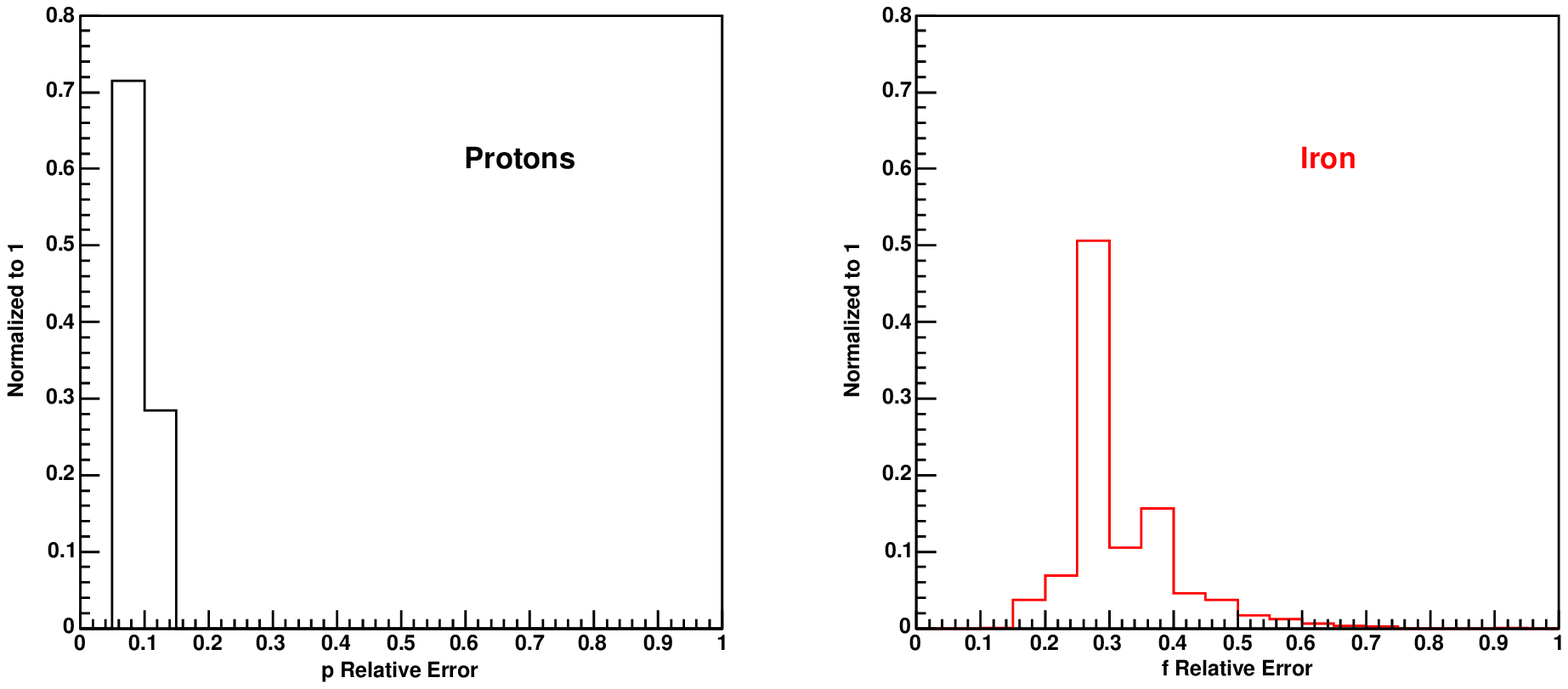}
 \caption{\label{fig:errordiff0.8} {\it Fractional uncertainty for $p$ (black) and $f$
 (red) calculated for simulated data sets with a parent sample comprised of 
 160 proton and 40 iron events.} }
\vskip 1 cm 
\end{figure}

We have tested the method with other mixtures of iron and protons and find 
that it gives reasonable rsults for all two-component mixtures.  
For instance, in Fig.\,\ref{fig:errordiff0.8} we 
find that the method still gives reasonable uncertainties in the case
where there is an 80:20 proton:iron mixture in  a 200 event parent distribution.  
The mode of the fractional error is $\sim 8\%$ ($\sim 28\%$) for protons (iron).  

\subsection{Eliminating Spectral Dependence in the Likelihood}

In practice, the energy spectra for protons and iron are unknown.
To include our lack of prior knowledge of the energy spectrum,
a slightly more sophisticated likelihood is used.

The method allows us to float $\alpha_p$ and $\alpha_f$.  
The likelihood can be maximized as a function
of $\alpha_p$ and $\alpha_f$ along with $p$ and $f$, but, due to limited 
statistics in the hypothesized simulated samples, we choose to 
marginalize these parameters.  
In principle, one would integrate over an infinite number of spectra. 
However, the CPU time for such a calculation is unrealistic, so here these spectra
will be marginalized using $E^{-2}$ and $E^{-3}$ spectra.  
Eq.\,(\ref{equation:bayes}) then becomes
\begin{equation}
\label{equation:bayes_modified}
P(p,f|D) = \frac{\sum_{\alpha_p=-2,-3}\sum_{\alpha_f=-2,-3}P(D|p,f,\alpha_p,\alpha_f)q(p,f,\alpha_p,\alpha_f)}{\sum_{p,f,\alpha_p,\alpha_f}P(D|p,f,\alpha_p,\alpha_f)q(p,f,\alpha_p,\alpha_f)}
\end{equation}
Other priors such as those concerning the interaction model can be introduced 
and marginalized in a similar manner.
Therefore, maximizing the probability $P(p,f|D)$
is to maximize the likelihood 
\begin{eqnarray}
l(D|p,f,\alpha_p,\alpha_f) \equiv P(D|p,f,\alpha_p,\alpha_f)q(p,f,\alpha_p,\alpha_f) \\\notag
=l(D|\{p_i\},\{f_i\},\epsilon_p,\epsilon_f).
\end{eqnarray}
or 

\begin{scriptsize}
\begin{equation}
\label{equation:likelihood_full}
l(D|\epsilon_p,\epsilon_f) = \sum_{\alpha_p=-2,-3}\sum_{\alpha_f=-2,-3}
\prod_{i=1}^{N~bins}
\int d\tilde{p}_i \int d\tilde{f}_i \left( \frac{\tilde{d}_i^{D_i}e^{-\tilde{d}_i}}{D_i!} \right)
\left( \frac{(\tilde{p}_i)^{P_{i}}e^{-\tilde{p}_i}}{P_{i}!} \right)
\left( \frac{(\tilde{f}_i)^{F_{i}}e^{-\tilde{f}_i}}{F_{i}!} \right).\\ 
\end{equation}
\end{scriptsize}

We can see that the introduction of priors does not change the distributions
in Figs.\,\ref{fig:diff} and \,\ref{fig:sigma} substantially
as seen by comparing the blue and green histograms.  
Its effect can be best seen by the upward shift in the mean relative 
uncertainty (between the blue and green histograms) in Fig.\,\ref{fig:errordiff_p}
and \ref{fig:errordiff_f}.

We conclude that the method gives an accurate prediction of the mean number of 
protons and iron.  For instance, with 100 proton and 100 iron events in a data 
sample with primary energies over $10^{19}$ eV, the uncertainty in $p$ and $f$ is $\sim 12\%$ 
with a better chance of obtaining a larger uncertainty.
This increase is expected as we have introduced some ``ignorance'' in the choice of spectra.
Further, the likelihood is able to accurately predict the $1\,\sigma$
uncertainties.

In the instance where we wish to sum over spectra, we would ideally integrate over 
every possible spectrum for protons and iron.  Computationally, this is obviously impossible.	
Instead, we compromise by summing over ``extremes'' of what we would expect for protons and iron.  
In this first approximation, we calculate
$L(D|\epsilon_p,\epsilon_f,\alpha_p=2,\alpha_f=2)$, $L(D|\epsilon_p,\epsilon_f,\alpha_p=2,\alpha_f=3)$, 
$L(D|\epsilon_p,\epsilon_f,\alpha_p=3,\alpha_f=2)$ and $L(D|\epsilon_p,\epsilon_f,\alpha_p=3,\alpha_f=3)$.  
That is, we calculate the likelihood with every combination of spectra for the proton and iron 
distributions.  Then, we sum these 4 likelihood and maximize the sum with respect to $\epsilon_p$ and $\epsilon_f$.  
Had we summed over generators (say QGSJET and SIBYLL), we would sum over 8 likelihoods where each likelihood is 
calculated with a different combination of spectra and hadronic generators.   
The calculation of uncertainties are made in the usual way with the summed likelihood.

\section{Discussion}
\label{section:conclusion} 
We  now investigate the case where the sample made of proton and iron
is ``contaminated'' with other components.  We start the discussion with 
gamma primaries.  AGASA has estimated the upper limit in the 
$\gamma$-ray flux to be $28\%$ for events above $10^{19}$ eV \cite{agasa_gammas}.  We therefore
create a 250 events sample with 100 protons, 100 iron, 50 gammas.  
The hypothesized distributions for gammas are based on 
1352 and 490 gammas with $E^{-2}$ and $E^{-3}$, respectively.
We use the gammas with an $E^{-3}$ spectrum to generate the parent distribution.
The result of the contamination is a large bias 
in $p$ and $f$ found in the black histogram in Fig.\,\ref{fig:diff}
and wrong uncertainties as seen in Figs.\,\ref{fig:sigma}, 
\ref{fig:errordiff_p} and \ref{fig:errordiff_f}.  However, if the likelihood is modified 
to 
\begin{scriptsize}
\begin{eqnarray}
\label{equation:likelihood_gammas}
l(D|\epsilon_p,\epsilon_f) =  \sum_{\alpha_p=-2,-3}\sum_{\alpha_f=-2,-3}
\prod_{i=1}^{N~bins} 
\int d\tilde{p}_i \int d\tilde{f}_i \\ \notag
\left( \frac{\tilde{d}_i^{D_i}e^{-\tilde{d}_i}}{D_i!} \right)
\left( \frac{(\tilde{p}_i)^{P_{i}}e^{-\tilde{p}_i}}{P_{i}!} \right)
\left( \frac{(\tilde{f}_i)^{F_{i}}e^{-\tilde{f}_i}}{F_{i}!} \right)
\left( \frac{(\tilde{g}_i)^{G_{i}}e^{-\tilde{g}_i}}{G_{i}!} \right)
\end{eqnarray}
\end{scriptsize}
where $\epsilon_g$, $G_{i}$, etc. carry the same definitions
for gammas as  $\epsilon_pp$, $P_{i}$, etc. carry for protons,
we see that the uncertainties once again make sense and the bias 
becomes negligible
as shown by the red histograms is Figs.\,\ref{fig:diff} and \ref{fig:sigma}.  In Figs.\,\ref{fig:errordiff_p} and\,\ref{fig:errordiff_f}  
we see an increase in the relative
uncertainties for iron while the proton uncertainties do not change significantly.  
An increase in the uncertainies would not be surprising as there is an overlap 
in the $X_{max}$ distributions between protons, iron and gammas.  
The lack of significant change is a result of the right combination
of  protons with an $E^{-2}$ spectrum and iron with an $E^{-2}$ spectrum
having a much larger likelihood value than the alternatives 
in the summation of $\alpha_p$ and $\alpha_f$.  
This demonstration shows that 
one must account for gamma-like primaries if there is a reasonable expectation 
that they may be in the data.

Accounting for all possible contributions from every imaginable primary 
is clearly impossible.  A number of questions then arise.  
For instance, is it sufficent to assume
a proton and iron mixture when analyzing a real data distribution?  
Is it necessary to assume, for instance, a proton, helium, carbon and iron 
mixture?  
Would such complex mixtures be more or less useful than simply quoting the
result assuming a proton and iron mixture?

These questions can be approached in two ways.  
If we had 
good reason to believe ultrahigh energy cosmic rays were composed 
exclusively of (say) protons, carbon and iron, then we would consider it
useful to measure the relative contributions from these three elements
regardless of their uncertainties.  
  
On the other hand, if we consider protons, carbon and 
iron as stand-ins for ``light,'' ``medium'' sized and ``heavy'' primaries,
then we can ask how sensitive the likelihood method is to these elements.
That is, if there were something resembling carbon mixed with protons and
iron, could it be detected to any accuracy?

\begin{figure}[t]
\includegraphics[width=1.\textwidth]{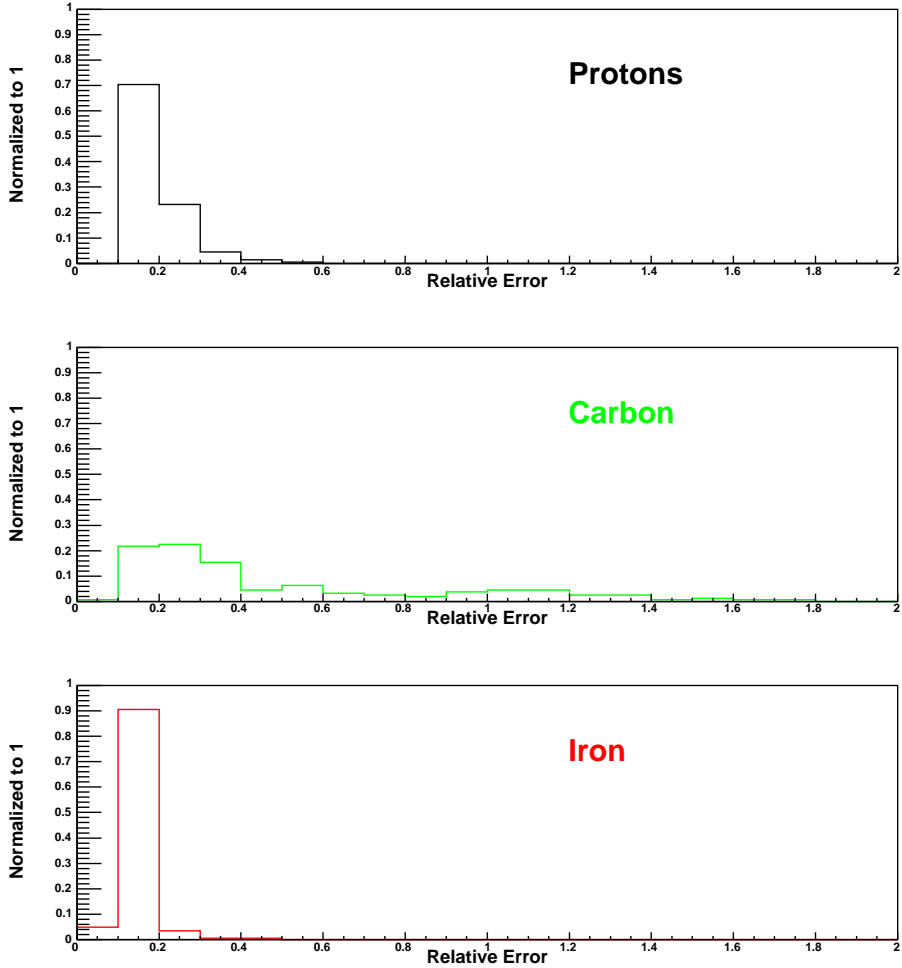}
 \caption{\label{fig:errordiff_carbon}  {\it Distribution of the relative errors for proton, carbon and 
 iron hypotheses evaluated using Eq.\,(\ref{equation:likelihood_gammas}) over simulated data sets 
 fluctuated from a parent sample of 100 proton, 50 carbon and 100 iron events.}}
\vskip 1 cm 
\end{figure}

In the latter vein, we build a ``fake'' data sample with 100 proton,
50 carbon and 100 iron primaries.  We then apply the full likelihood
in Eq.\,(\ref{equation:likelihood_gammas}), replacing the gamma term with carbon,
and evaluate the distribution of uncertainties (Fig.\,\ref{fig:errordiff_carbon}).
Although we find the uncertainties are meaningful,
the large uncertainties for carbon indicate that the resolution 
is not sufficient to make statements about the fraction 
of medium-sized primaries
in this type of data sample by inserting the carbon hypothesis.
From this perspective, it is less important to account for carbon primaries than gamma 
primaries.

We also study the case where two hypothesized elements are indistinguishable.  
We create a fake data set of 250 events with 150 proton and 100 iron events.  
A proton distribution, a second identical proton distribution and 
an iron distribution act as hypotheses.  
That is, the likelihood is calculated using Eq.\,(\ref{equation:likelihood_gammas})
substituting a second proton distribution for the gamma distribution.  
The algorithm that maximizes the likelihood, unable to distinguish how many 
of the 150 proton are attributable
to the first proton distribution as opposed to the second, settles on an arbitrary 
number of proton events between 0 and $\sim 150$ for the first proton hypothesis ($p_1$).  
What remains of $\sim 150$ events is attributed to the second proton hypothesis ($p_2$). 
The uncertainties are then calculated for $p_1$ and $p_2$
by fixing the number of events for $p_1$ ($p_2$) fluctuating 
$p_2$ ($p_1$) $\pm\,1\sigma$ according to Eq.\,(\ref{equation:error}).  
As there are large correlations between $p_1$ and $p_2$ that are not considered
when calculating the uncertainties in 
this way, the method underestimates the uncertainties 
($1\,\sigma$ corresponds to $15\%$ confidence region).
Therefore, before an element is inserted as a hypothesis, the correlations
between the various distributions need to be understood or one needs to verify that
the distributions
are sufficiently uncorrelated such that the results have meaningful uncertainties.  

\begin{figure}[t]
\includegraphics[width=1.\textwidth]{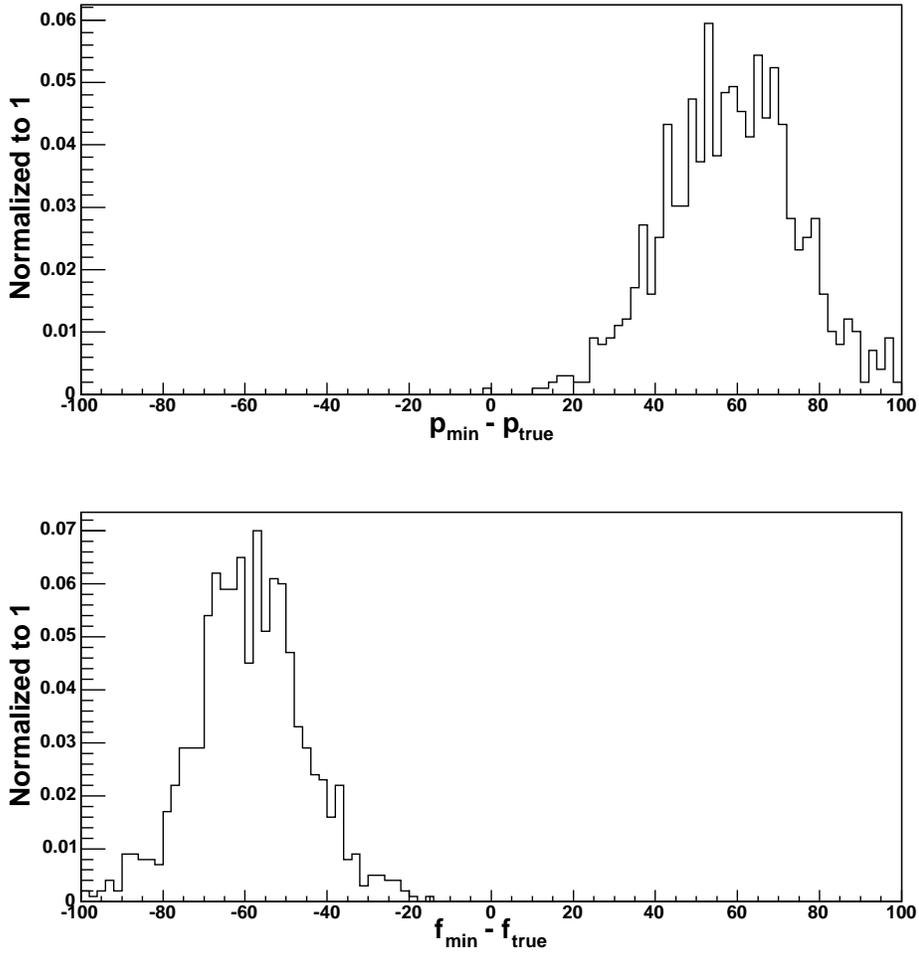}
 \caption{{\it \label{fig:diff_sibyll} Histograms of $p$-$p_{true}$ (top) and $f$-$f_{true}$ (bottom)
 for $p_{true}=100$ and $f_{true}=100$ where the simulated data is generated with SIBYLL
 and the likelihood calculation uses proton and iron events generated with QGSJET.  }}
\vskip 1 cm 
\end{figure}

Also, one may consider other hadronic models.  In principle, the 
hadronic models can be marginalized by simply ``summing'' over 
them.  $X_{max}$ distributions for protons, iron and gammas
with $E^{-2}$ and $E^{-3}$ spectra would be generated using QGSJET and 
SIBYLL hadronic models.  The complexity of the likelihood
and the amount of Monte Carlo increases sharply with the number of priors.
For the sake of time and simplicity the full calculation 
is not made here.  However, to show that it is important to
consider different hadronic models, we generate a simulated 
data set with 100 proton and 100 iron events with the SIBYLL
hadronic model.  We then fluctate this distribution many times
in the same fashion described above each time calculating
the maximum likelihood using proton and iron events generated with 
QGSJET.  
Figure\,\ref{fig:diff_sibyll} shows the $\sim 60 \%$ shift in the mean
number of protons and iron resulting from 
calculating the likelihood using QGSJET.   
Regardless, this shift shows that the
composition fractions cannot be properly calculated without considering at least
the QGSJET and SIBYLL hadronic models.

We have discussed widening the measurement to include such knowledge
and given an example of how one might introduce prior knowledge 
by introducing and then marginalizing the unknown spectral dependence of iron and protons.
Other
parameters concerning the 
extensive air shower and atmospheric uncertainties
can be introduced and marginalized provided sufficient data and CPU time.
In so doing, uncertainties are rigorously compounded in the calculation.  

Further, we have investigated the consequence of contaminating the simulated
data with gammas and carbon.  Proton, iron and gamma hypotheses
are sufficient if one is only interested in stand-ins for ``heavy,'' ``light,'' 
and ``lighter'' elements.  However, it is important to
consider all elements that one reasonably expects to be in the data.

This method of measuring the composition is different from 
the method using elongation rate.  It uses the full distribution of $X_{max}$ to determine the composition 
as opposed to the elongation rate which uses the mean.
Also, the measured elongation is independent of any model assumptions
although its interpretation requires one to compare the measurement 
to what is expected for various compositions.  The method is currently being
applied to HiRes stereo events and results will be presented in a separate paper
where the composition will be evaluated for events in intervals
of $10^{18.5}$\,eV, $10^{19.0}$\,eV,  $10^{19.5}$\,eV, etc.



\begin{thebibliography}{99}
%
\bibitem{agasa_comp} 
N. Hayashida {\it et al.}, J. Phys. G 21\,(1995)\,1101.
%
\bibitem{bird1993}  
D.J. Bird {\it et al.}, Phys. Rev. Lett. 71\,(1993)\,3401.
%
\bibitem{proto} 
T. Abu-Zayyad {\it et al.}, Phys. Rev. Lett. 84\,(2000)\,4276.
%
\bibitem{archbold}  
R.U. Abbasi {\it et al.}, Astrophys. J. 622 (2005) 910-926.
%
\bibitem{bhat}
P. Bhat, H. Prosper, and S. Snyder, Phys. Lett. B 407\,(1997)\,73.
%
\bibitem{101}  C. Song, Study of Ultra High Energy Cosmic Rays with the High Resolution Fly's Eye Prototype Detector\,(2001)\,(New York: Columbia University)
%
\bibitem{89}  D. Heck {\it et al.}, University of Kalsruhe Report No. FZKA-6019\,(1998)\,(Hamburg: University of Kalsruhe)
%
\bibitem{87}  N.N. Kalmykov, S.S. Ostapchenko, and A.I. Pavlov, Nucl. Phys. B. (Proc. Suppl.), 52B\,(1997)\,17
%
\bibitem{102} T. Abu-Zayyad {\it et al.}, Astropart. Phys., 16\,(2001)\,1
%
\bibitem{nim2002}
J. Boyer {\it et al.}, Nucl. Instr. Meth. A 482\,(2002)\,457.
%
\bibitem{star2002}
P.A. Sadowski {\it et al.}, Astropart. Phys. 18\,(2002)\,237.
%
\bibitem{matthews2003}
J.N. Matthews {\it et al.}, Proc. of 28th ICRC, Tsukuba, Japan, 350 (2003).
%
\bibitem{ref9}
R.U. Abbasi et al., Astrophys. J. 610 (2004) L73.
%
\bibitem{jaynes}
E.T. Jaynes and G. Larry Bretthorst (Ed.), {\it Probability Theory: The Logic of Science},
Cambridge, 2003.
%
\bibitem{powell}
M.J.D. Powell, Comput. J. 7\,(1964)\,255.
%
\bibitem{agasa_gammas}
K. Shinozaki {\it et al.}, Astrophys. J.  571\,(2002)\,L117.

\end{thebibliography}
\end{document}